%% file: SIR_manuscript_arxiv.tex
\documentclass[12pt]{article}
\usepackage{natbib}
\usepackage{hyperref}
\usepackage{grffile}
\usepackage{graphicx}
\usepackage{subcaption}
\usepackage{amssymb,amsmath,amsthm}
\usepackage{xcolor}
\usepackage{xspace}
\usepackage[nameinlink,capitalize]{cleveref}
\usepackage{cleveref}
\usepackage[margin=1in]{geometry}
\usepackage{lineno}
\usepackage{pdflscape}
\usepackage{enumerate}
\usepackage{authblk}

\input{modeldefs.tex}

\usepackage{array}
\newcolumntype{L}[1]{>{\raggedright\let\newline\\\arraybackslash\hspace{0pt}}m{#1}}
\newcolumntype{C}[1]{>{\centering\let\newline\\\arraybackslash\hspace{0pt}}m{#1}}
\newcolumntype{R}[1]{>{\raggedleft\let\newline\\\arraybackslash\hspace{0pt}}m{#1}}

\usepackage{xspace}

\newcommand{\fref}[1]{Fig.~\ref{#1}}
\newcommand{\appref}[1]{Appendix~\ref{app:#1}}

\newcommand{\percap}{\emph{per capita}\xspace}
\newcommand{\Rnum}{\ensuremath{\mathcal{R}_0}\xspace}
\newcommand{\covid}{COVID-19\xspace}
\newcommand{\pro}[1]{\ensuremath{\frac{\partial #1}{\partial \rho}}}
\newcommand{\pder}[2]{\ensuremath{\frac{\partial#1}{\partial#2}}} 
\newcommand{\testing}[1]{\ensuremath{{\cal T}_{#1}}\xspace}
\newcommand{\testinghat}[1]{\ensuremath{\hat{\cal T}_{#1}}\xspace}
\newcommand{\testtarget}{\ensuremath{w\_{IS}}}
\newcommand{\oldtesttarget}{\ensuremath{\frac{w_I}{w_S}}\xspace}
\newcommand*\subtxt[1]{_{\textnormal{#1}}}
\DeclareRobustCommand\_{\ifmmode\expandafter\subtxt\else\textunderscore\fi}

\newcommand{\nocomment}[3]{}

\usepackage{ragged2e}
\usepackage{array} 

\theoremstyle{definition} 

\title{Testing and Isolation Efficacy:\\Insights from a Simple Epidemic Model}
\author[1]{Ali Gharouni}
\author[1]{F.M. Abdelmalek}
\author[1,3,5]{David J. D. Earn}
\author[2,4,3]{Jonathan Dushoff}
\author[1,2,3]{Benjamin M. Bolker}

\affil[1]{Department of Mathematics \& Statistics, McMaster University, Hamilton, Canada}
\affil[2]{Department of Biology, McMaster University, Hamilton, Canada}
\affil[3]{Michael G. DeGroote Institute for Infectious Disease Research, McMaster University, Hamilton, Canada}
\affil[4]{South African Centre for Epidemiological Modelling and Analysis, University of Stellenbosch, Stellenbosch, South Africa}

\begin{document}
\maketitle


\section{Abstract}

Testing individuals for pathogens can affect the spread of epidemics. 
Understanding how individual-level processes of sampling and reporting test results can affect community- or population-level spread is a dynamical modeling question.
The effect of testing processes on epidemic dynamics depends on factors underlying implementation, particularly testing intensity and on whom testing is focused.
Here, we use a simple model to explore how the individual-level effects of testing might directly impact population-level spread.
Our model development was motivated by the \covid epidemic, but has generic epidemiological and testing structures. 
To the classic SIR framework we have added a \percap testing intensity, and compartment-specific testing weights, which can be adjusted to reflect different testing emphases --- surveillance, diagnosis, or control. 
We derive an analytic expression for the relative reduction in the basic reproductive number due to testing, test-reporting and related isolation behaviours.
Intensive testing and fast test reporting are expected to be beneficial at the community level because they can provide a rapid assessment of the situation, identify hot spots, and may enable rapid contact-tracing. 
Direct effects of fast testing at the individual level are less clear, and may depend on how individuals' behaviour is affected by testing information.
Our simple model shows that under some circumstances both increased testing intensity and faster test reporting can \emph{reduce} the effectiveness of control, and allows us to explore the conditions under which this occurs.
Conversely, we find that focusing testing on infected individuals always acts to increase effectiveness of control.

\section{Introduction}

The observed dynamics of the \covid epidemic have been driven both by epidemiological processes (infection and recovery) and by testing processes (testing and test reporting). In addition to shaping epidemic observations (via case reports), testing processes also alter epidemiological dynamics \citep{peto2020covid,taipale2020population}. Because individuals with confirmed infections (positive tests) are likely to self-isolate, and individuals who are awaiting the results of a test may also do so, testing will generally increase the number of people who are isolating and hence reduce epidemic growth rates. We developed a mechanistic model that incorporates epidemic processes and testing in order to explore the effects of testing and isolation on epidemic dynamics.

If testing influences behaviour, then epidemic dynamics will depend on who gets tested.
The impacts of testing will depend both on testing intensity (tests performed per day) and on how strongly testing is focused on people who are infectious.
This level of focus depends in turn on the purpose and design of testing programs. 
When testing is done for the purposes of disease surveillance \citep{foddai2020base}
tests are typically conducted randomly (or using a stratified random design) across the population in order to make an unbiased assessment of population prevalence.

Over the course of the \covid pandemic, however, the vast majority of testing has been done with other goals --
primarily diagnostic (determining infection status for clinical purposes) \citep{phua2020intensive,who2020global}, or for control (determining  infection status in order to isolate cases that have been found by contact tracing) \citep{aleta2020modelling,kucharski2020effectiveness,grassly2020comparison,smith2020adherence}, which we characterize as \emph{targeted} testing strategies.
In these situations, testing probabilities can differ sharply across epidemiological compartments; in our dynamical model, we will characterize these probabilities by assigning a testing weight to each compartment that determines the \emph{relative} probability that an individual in that compartment will be selected for testing (see Methods). 

Diagnostic testing focuses on people with infection-like symptoms; thus the relative testing weights for infected people will depend on the relative probability of infected people having symptoms. For \covid infection, the testing weights will depend on the proportion of asymptomatic infections, the time spent pre-symptomatic vs.\ symptomatic during the course of an infection, and on the incidence of \covid-like symptoms among people in the population \emph{not} infected with \covid. Testing for epidemic control focuses on known contacts of infected people; in this case the testing weights for infected vs.\ uninfected people will depend on the probability of infection given contact, as well as the effectiveness of the system for identifying suspicious contacts.

When a new infectious disease emerges, it is important to determine whether it will grow exponentially in a susceptible population, and if so at what rate $r$ \citep{Ma+14}.  The condition for positive exponential growth ($r>0$) is commonly expressed as $\Rnum>1$, where the basic reproduction number $\Rnum$ is the expected number of secondary infections arising from a typical infective individual in a completely susceptible population \citep{dietz1993estimation}.  Although the value of $\Rnum$ cannot completely characterize the dynamics of our model \citep{shaw2021what}, it does give a simple and widely accepted index for the difficulty of control, as well as an indication of the likely final size of an epidemic \citep{ma2006generality,miller2012note}.

In order to understand the effect of testing processes on epidemic dynamics, we expanded one of the simplest mechanistic epidemic models---the standard deterministic SIR model\citep{KermMcKe27,AndeMay91}---to include testing components. This model provides a sensible platform to link the modeling of epidemic and testing components and study their interaction. We studied the effects of testing intensity, rate of test return, and isolation efficacy, on transmission probability and epidemic dynamics when different levels of testing focus (from random to highly targeted) are in place.

\section{Methods}

\begin{figure}
\begin{center} 
\includegraphics[scale=1.5]{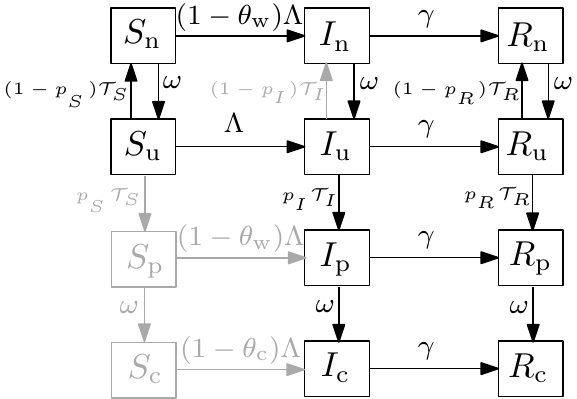}
\caption{\small Flowchart of the SIR (Susceptible-Infectious-Recovered) model, \ref{model}. The disease-based status of a compartment $X$ ($X \in \{S,I,R\}$) is combined with the testing status including $X\_u$, $X\_p$, $X\_n$ and $X\_c$, for \emph{untested}, waiting-for-\emph{positive}, waiting-for-\emph{negative}, or \emph{confirmed positive}, respectively. The force of infection is denoted by $\Lambda$ (Eq.~\eqref{Lambda}); $\gamma$ is the recovery rate; $\omega$ is the rate of test return; and \testing{X} (Eq. \eqref{test}) and $p_X$ represent the \percap testing rate and the sensitivity (probability of positive tests), respectively, for compartment $X$. For further description of the parameters see \cref{tab:params}.
Note that there is a slight mismatch in the top-to-bottom order of the testing-based compartments of each disease-based compartment $X$ between this flowchart and the model equations \eqref{model}; here we have switched  $X\_u$ and $X\_n$ for visual clarity.
\label{fig:flowchart}}
\end{center} 
\end{figure}

Our model groups individuals based on disease status (Susceptible, Infectious or Recovered) and testing status (\emph{untested}, waiting-for-\emph{positive}, waiting-for-\emph{negative}, or \emph{confirmed positive}) (\fref{fig:flowchart}).  The testing status of an individual in a given disease compartment $X$ (where $X \in \{S,I,R\}$) is denoted by a subscript, namely $X\_u$, $X\_p$, $X\_n$ and $X\_c$, for \emph{untested}, waiting-for-\emph{positive}, waiting-for-\emph{negative}, or \emph{confirmed positive}, respectively.  Two `accumulator' compartments, $N$ and $P$, are included in order to collect cumulative reported negative or positive tests. The model equations \eqref{model} and details of calculation of the basic reproduction number $\Rnum$ are presented in \appref{R0}.

Table~\ref{tab:params} defines the model parameters, which are generally \percap flows between compartments, or modifiers to these flow rates. The novel component of the model lies in the compartment-specific relative testing weights $w_S$, $w_I$ and $w_R$; these give the relative rates at which people in the $S$, $I$, and $R$ compartments are tested, respectively. Thus, we can specify different levels of testing focus from random (all weights equal) to highly targeted (higher weights in more intensively tested compartments). For example, $w_I/w_S=3$ means that infected individuals are tested at three times the \percap rate of susceptible individuals. 

In order to allow parameterization of the model by the total (overall) \percap testing rate, we define the weighted size of the testing pool $W = w_S S\_u + w_I I\_u + w_R R\_u$, and calculate a scaling parameter for testing as:
\begin{equation}
\label{sigma}
\sigma = \frac{\rho N}{W},
\end{equation}
where $\rho$ is the \percap testing intensity for the population, defined as the number of daily tests administered in a population of size $N$.
Thus, the \percap testing rate for compartment $X \in \{S,I,R\}$ is 
\begin{equation}
\label{test}
\testing{X}=\sigma w_X.
\end{equation}
For a highly sensitive test, infected people typically flow through to the ``confirmed positive'' ($I\_c$, $R\_c$) compartments and are thus not considered for further testing. Over the course of the epidemic, a sufficiently large fixed testing rate as specified in \eqref{sigma} can exhaust the pool of people available for testing, leading to a singularity when too few people are left untested to support the specified rate. Although this phenomenon does not affect our analysis of $\Rnum$, it can affect model dynamics (we present an adjustment to the model that solves this problem in \appref{singularity}).

The classical SIR model assumes a well-mixed population; homogeneity of the population (i.e., all individuals are equally susceptible and equally infectious with the same recovery rate when infected); exponentially distributed duration of infection; and large population size \citep{keeling2011modeling}. In addition to these standard assumptions, our model assumes: 
\begin{enumerate}[(i)]
\item there is a single force of infection (new infections per unit time per susceptible), $\Lambda$, defined as
  \begin{equation}
  \label{Lambda}
  \Lambda=\frac{\beta}{N} \big(I\_u + (1-\theta\_w)(I\_n+I\_p) + (1-\theta\_c)I\_c \big),
  \end{equation}
with transmission rate $\beta$; $\theta\_w$ is the isolation efficacy (reduction of the probability of transmission) for individuals waiting for test results, while $\theta\_c$ is the isolation efficacy for individuals who have received a ``confirmed'' positive test (Table~\ref{tab:params}). Susceptible individuals who are waiting for test results experience an additional transmission reduction factor of $1-\theta\_w$ (\fref{fig:flowchart}). 
\item confirmed-positive individuals isolate at least as effectively as those awaiting test results, i.e.,
$$0\leq\theta\_w\leq\theta\_c\leq 1.$$ 
\end{enumerate}
For simplicity we assume that tests are perfectly \emph{specific} --- uninfected individuals never test positive ($p_s=0$). Thus there are no waiting-for-positive or confirmed-positive susceptible individuals, which reduces the number of model states from 12 to 10.

\begin{table}[htp]
\centering
{\RaggedRight
\begin{tabular}{ c | m{6cm} | c | c }
  \textbf{Symbol} & \textbf{Description} & \textbf{Unit} & \textbf{Value} \\ \hline
  $N$     & Total population size & people & $10^6$ \\ \hline
  $\omega$  & Rate of test return, i.e., rate of onward flow from ``waiting'' to ``confirmed'' or ``untested'' compartments  & 1/day & - \\ \hline
  $\gamma$ & Recovery rate & 1/day & 1/6 \\ \hline 
  $\rho$   & \percap testing intensity & 1/day & - \\ \hline 
  $\theta\_w$ & Isolation efficacy (reduction of the transmission probability) for ``waiting'' individuals & - & - \\ \hline
  $\theta\_c$ & Isolation efficacy for ``confirmed positive'' individuals & - & -  \\ \hline
  $\beta$ & Transmission rate & 1/day & 0.5 \\ \hline
  $\Lambda$ & Force of infection & 1/day & - \\ \hline
  $p_S$ & Probability of positive tests for $S$ ($= 1-\textrm{specificity}$) & - & 0 \\ \hline
  $p_I$ & Probability of positive tests for $I$ ($= \textrm{sensitivity}$) & - & 1 \\ \hline
  $p_R$ & Probability of positive tests for $R$ ($= 1-\textrm{specificity}$) & - & 0.5 \\ \hline
  $w_S, w_I, w_R$ & Relative testing weights & - &
  \begin{minipage}[t]{0.21\columnwidth}%
    Random:~$\{1,1,1\}$ \\ Targeted:~$\{0.3,1,1\}$
  \end{minipage} \\
\end{tabular}
} 
\caption{\label{tab:params}Parameters of the model specified  by the flow chart in \fref{fig:flowchart} and equations~\eqref{model}.}
\end{table}

The Disease-Free Equilibrium (DFE) for the expanded SIR model (Eqs.~\ref{model}) is found by setting the infected compartments to 0 and solving for the unknowns. The DFE is
\begin{equation}
\label{dfe}
S\_n^*= \frac{\rho}{\omega} N, \ S\_u^*= N-S_n^*, \text{~and~} I_j=R_j=0 \ \text{for all }j.
\end{equation}
The corresponding \percap testing rate (Eq.~\ref{test}) for the infected compartment $I$ at DFE is one of the key analysis parameters and can be simplified as 
\begin{equation}
\label{eq:fi}
\testinghat{I} = (\omega\rho/(\omega-\rho))w_I/w_S \quad .
\end{equation}

The basic reproduction number, $\Rnum$, was calculated by using the next-generation matrix method \citep{van2002reproduction}. We write $\Rnum$ as
\begin{equation}
\label{R0}
\Rnum= \frac{\beta}{\gamma} \left(1-\Delta\right), 
\end{equation}
where $\beta/\gamma$ is the classical value for a simple model \citep{keeling2011modeling}, and $1-\Delta$ is the proportional reduction due to testing and isolation processes. $\Delta$ therefore measures the ``effectiveness of control'': how much these processes in reducing spread from low levels, and is in turn given by:
\begin{equation}
  \label{eq:del4a}
  \Delta= \frac{1}{C N}\big(C_1 S\_u^*+(C_2(1-\theta\_w)+C \theta\_w)S\_n^*\big),
\end{equation}
where
\begin{align}
\label{eq:C12}
C_{\phantom 1} &= (\omega+\gamma) \Big(\gamma(\omega+\gamma)+(\gamma+\omega p_I)\testinghat{I}\Big),\\
C_1 &= (\omega+\gamma)(\theta\_w \gamma+\theta\_c \omega p_I) \testinghat{I},\\
C_2 &= \Big( \omega\gamma(1+p_I)\testinghat{I}+\gamma^2(\omega+\gamma+\testinghat{I})\Big)\theta\_w + \omega^2p_I \testinghat{I} \theta\_c.
\end{align}
(\appref{R0} gives a detailed derivation of these expressions.)
This explicit formula enables us to study the effects of testing and isolation parameters on $\Rnum$ both analytically and via numerical solutions.
We are specifically interested in parameters that could be manipulated by public health policy: isolation efficacy, $\theta\_c$ and $\theta\_w$; \percap testing intensity, $\rho$; and the rate of test return, $\omega$. In particular, we look at the partial derivatives of $\Delta$ with respect to these parameters (Appendices~\ref{app:rho} and \ref{app:omega}). 
We derived general expressions for these derivatives. However, we analyzed the effect of $\omega$ on $\Delta$ for the special case of low testing intensity. Specifically, by making the restriction $\rho \ll 1$, we are able to Taylor-expand $\Delta$ at $\rho=0$, use the linear approximation with respect to $\rho$ and analyze the resulting simplified derivatives to illustrate a surprising non-monotonic relationship between $\Delta$ and $\omega$. 

Analytic calculation of the next-generation matrix and simplification of the $\Rnum$ expression, were performed in Maple\textsuperscript{\texttrademark} \citep{maple14}; numerical calculation and contour plots were done in R \citep{r}.
We computed the values and contours of $\Delta$ at both low (\fref{pan}) and high (\fref{pan2}) testing intensities, and for both random testing ($w_S=w_I=w_R=1$) and targeted testing ($w_S=0.3$; $w_I=w_R=1$). Because it is expressed as a proportion of $\Rnum$, the effectiveness of control $\Delta$ is (at least in the $\rho \ll 1$ case, Eq.~\ref{eq:lin}) independent of the transmission rate $\beta$, and hence of $\Rnum$ in the case where we vary $\Rnum$ by changing the transmission rate for a fixed generation interval.

The low-testing case (\fref{pan}) reflects the case where testing intensity $\rho$ is small relative to the population size. Specifically, $\rho \in [0,0.013]$, and test return rate $\omega\in [1/12,2]$. This testing intensity is of the correct order of magnitude (although typically larger than) testing rates during the \covid pandemic, i.e., a maximum of 1.3\% of the population per day (approximately four times the maximum testing rate in Ontario, Canada in mid-2021). The less realistic high-testing case (\fref{pan2}) is included to highlight the occurrence of non-monotonic changes in $\Rnum$ with respect to $\rho$.
In \fref{pan2} the maximum capacity of $\rho$ is larger relative to the population size, $\rho \in [0,1/5)$ and the test return rate $\omega\in [1/5,2]$; these values are clearly unrealistic for a large population but might be relevant for small populations undergoing focused testing, such as a sports league or university. In these figures, the implied baseline reproduction number (for the SIR model without testing) is $\Rnum=\frac{\beta}{\gamma}=3$.  The different ranges of test return rates $\omega$ for the cases of low and high testing intensities is due to the restriction $\rho<\omega$, which is a requirement for a feasible DFE (\ref{dfe}).
  
\section{Results}

We presented $\Rnum$ as the product of the classical reproduction number, $\beta/\gamma$, and the proportional reduction due to testing and isolation, $1-\Delta$, \eqref{R0}.
We can use the formula for $\Delta$ \eqref{eq:del4a} to make a number of straightforward inferences about parameters that affect $\Rnum$ monotonically, i.e., for which the associated partial derivative of $\Delta$ always has the same sign (see Appendices).

\begin{enumerate}

\item \label{p1:eta} Increasing isolation efficacy for waiting ($\theta\_w$) and confirmed-positive ($\theta\_c$) individuals always increases $\Delta$ (Eqs.~\ref{eq:del4_theta},~\ref{eq:d_del4_thetac},~\ref{eq:d_del4_thetaw});
\item \label{p1:rho} Higher testing intensity $\rho$ increases $\Delta$ if
testing is random (all $w_X$ equal) or testing intensity ($\rho$) is small (Eq.~\ref{eq:dd3dr}).
\item \label{p1:omega} Increasing the rate of test return ($\omega$) always increases $\Delta$ if waiting individuals do not isolate ($\theta\_w=0$) (Eq.~\ref{eq:dlindo}).
\item \label{p1:w} Increasing testing focus, i.e., changing the testing weights from random ($w_S=w_I$) toward targeted  ($w_S<w_I$), always increases $\Delta$ (Eq.~\ref{eq:d_del_wis}).
\end{enumerate}

\newpage
\begin{figure}[h!]
\centering
\begin{subfigure}[t]{.49\textwidth}
\centering
\includegraphics[width=\linewidth]{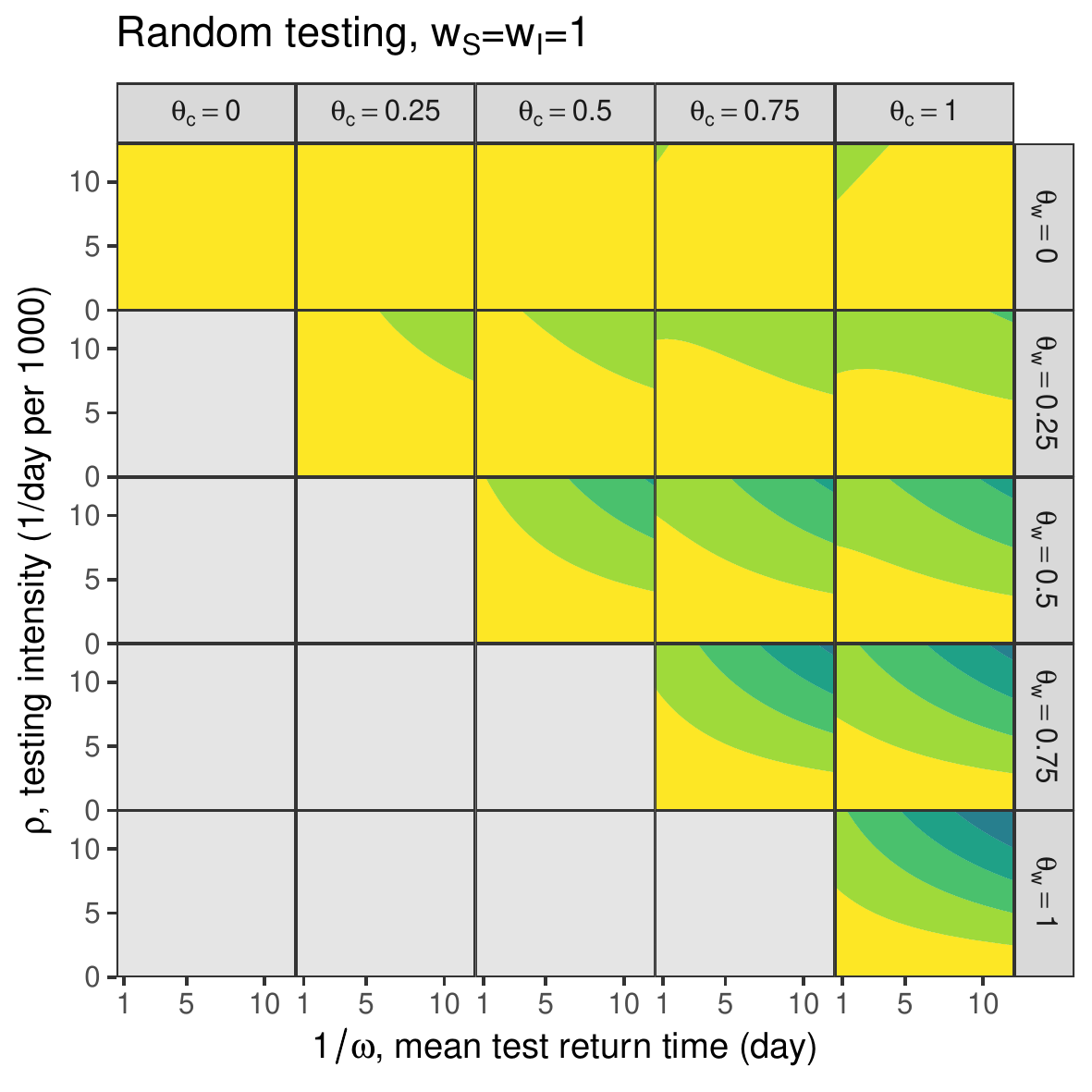}
\caption{}\label{p.a}
\end{subfigure}
\begin{subfigure}[t]{.49\textwidth}
\centering
\includegraphics[width=\linewidth]{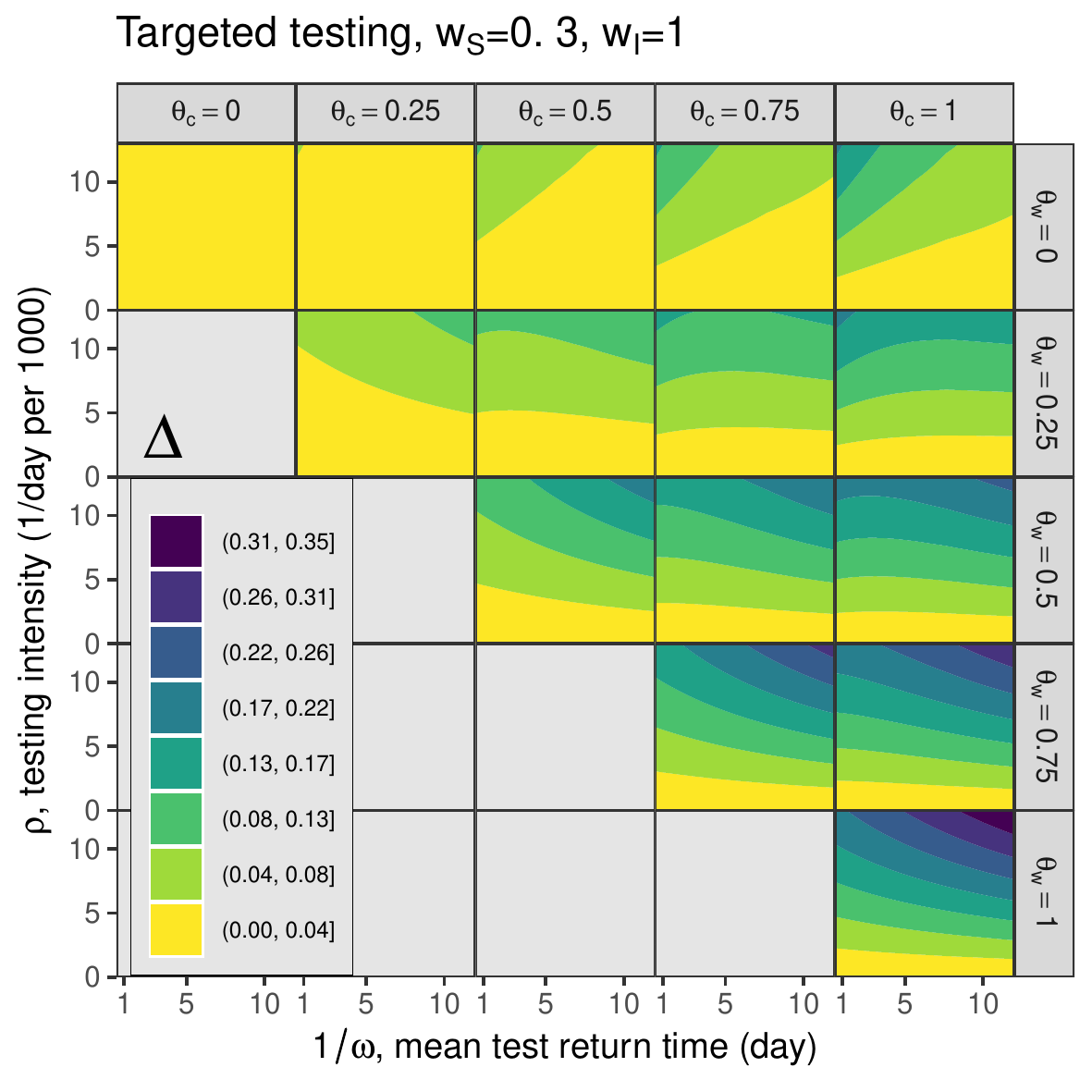}
\caption{}\label{p.b}
\end{subfigure}
\caption{
{\bf Effectiveness of testing and isolation in reducing $\Rnum$ at low \percap testing intensity ($\rho$).}
Numerical evaluation of the effectiveness of control ($\Delta$: eq.~\ref{eq:del4a}), over a range of testing and isolation parameters. Parameter values (Table~\ref{tab:params}):
$\beta=\betaparam/$day, $1/\gamma= \invgammaparam$ days (baseline $\Rnum=\Rnumparam$, $r=\rparam$); $\omega \in [1/12,2]/$day;  $\rho \in [0,0.013]/$day per capita; $\theta\_w$ and $\theta\_c$ vary between 0 (no effect of isolation) and 1 (complete elimination of transmission); $p_S=0$, $p_I=1$ and $p_R=0.5$. Only parameter sets where $\theta\_c \geq \theta\_w$ (confirmed-positive individuals isolate more effectively than waiting individuals) are shown; the alternative case, $\theta\_w > \theta\_c$, is unrealistic. Contours of $\Delta$ are plotted for (a) random testing ($w_S=w_I=w_R=1$) and (b) targeted testing ($w_S=0.3$; $w_I=w_R=1$). 
}
\label{pan}
\end{figure}

\begin{figure}[h!]
\centering
\begin{subfigure}[t]{.49\textwidth}
\centering
\includegraphics[width=\linewidth]{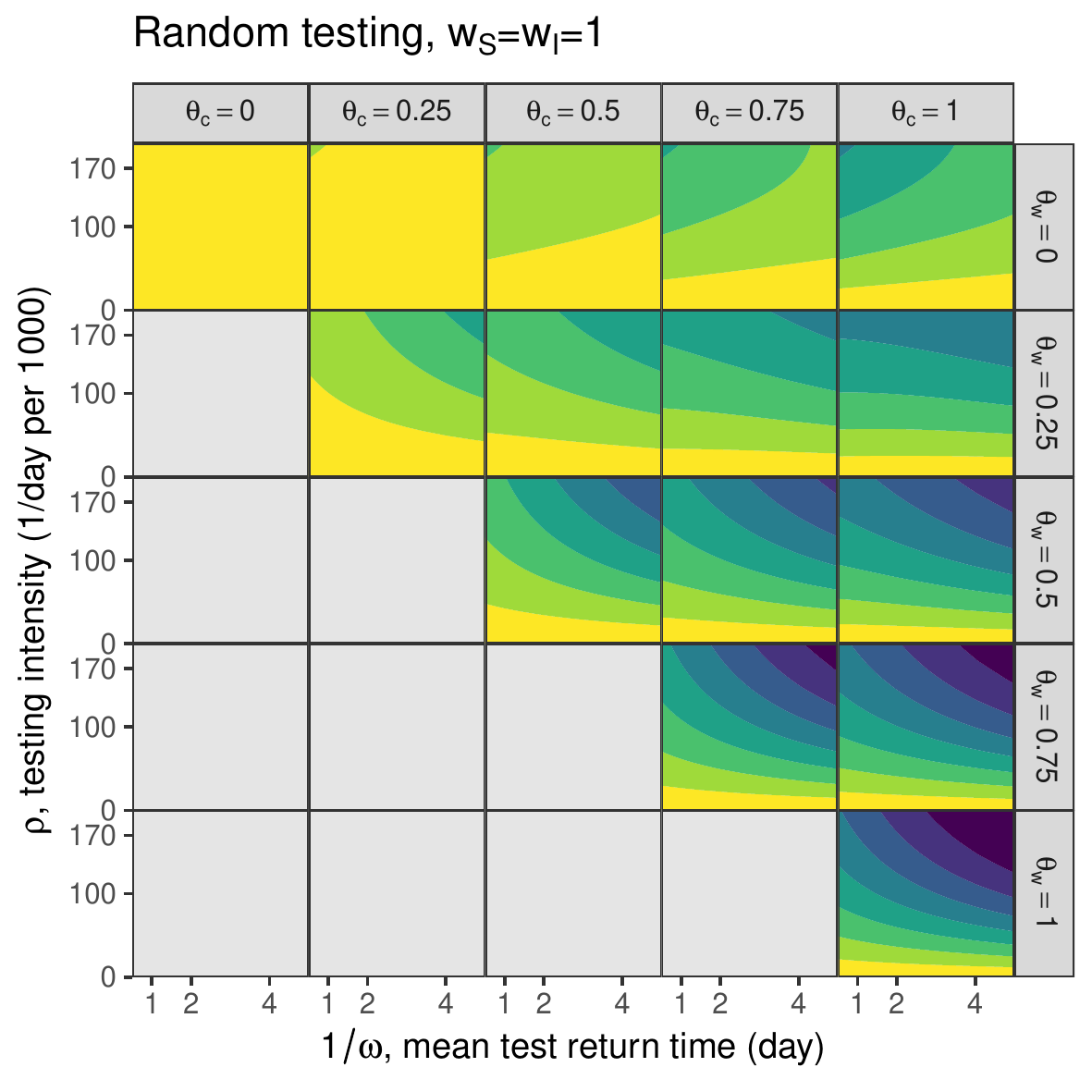}
\caption{}
\end{subfigure}
\begin{subfigure}[t]{.49\textwidth}
\centering
\includegraphics[width=\linewidth]{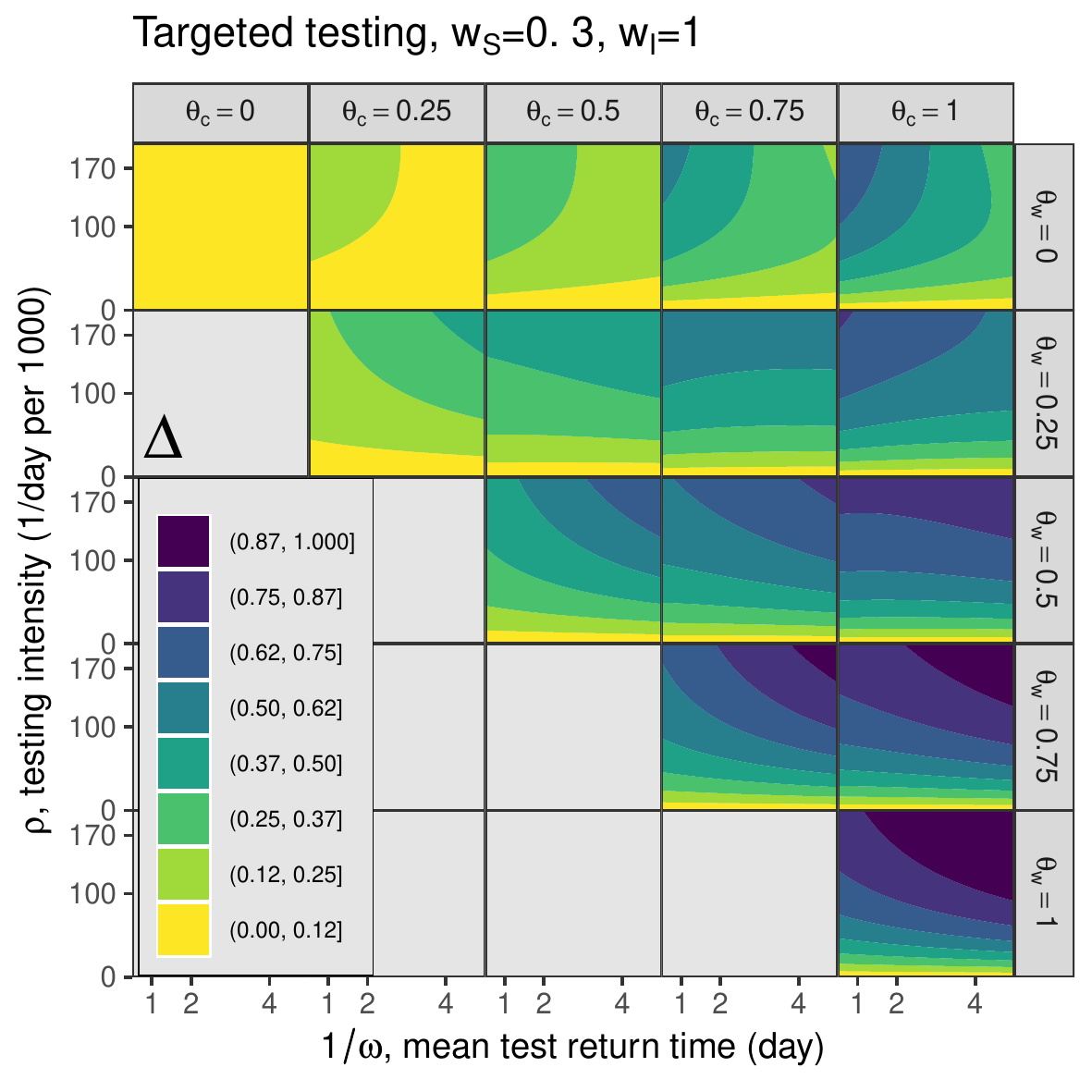}
\caption{}
\end{subfigure}
\caption{
  {\bf Effectiveness of testing and isolation in reducing $\Rnum$ at high \percap testing intensity.}
  Numerical evaluation of the effectiveness of control ($\Delta$: eq.~\ref{eq:del4a}), over a range of testing and isolation parameters. Parameters as in \fref{pan} except: $\omega \in [1/5,2]/$day, $\rho \in [0,1/5)/$day. As in \fref{pan}, subplots show (a) random testing where $w_S=w_I=w_R=1$ and (b) targeted testing where $w_S=0.3$ and $w_I=w_R=1$.
}
\label{pan2}
\end{figure}

However, there are also two specific cases where $\Delta$ changes non-monotonically, in counterintuitive directions, as a function of testing and isolation parameters.

\begin{itemize}
\item We would generally expect increasing testing delays to increase \Rnum, thus decreasing effectiveness of control $\Delta$. This is in fact what happens when waiting individuals do not isolate ($\theta\_w =0$, top row of \fref{pan}) --- as we move to the right within each plot in this row, $\Delta$ decreases.
However, when waiting individuals isolate ($\theta\_w>0$), we more often see the opposite effect: longer testing delays lead to a greater control effect $\Delta$ (reduced $\Rnum$). The reason is that people waiting for negative tests are assumed to continue to isolate; this applies both to susceptibles and to people who became infected while waiting for negative test results. This effect outweighs the effect of confirmed individuals isolating, except when this isolation parameter ($\theta\_c$) is substantially greater than $\theta\_w$. This result depends on the idea that, all else equal, people who have to wait longer for test results isolate at the same level (but for a longer time) as they would if the wait were shorter.
\item \fref{pan} also shows that greater testing intensity (increasing $\rho$) generally increases the effectiveness of control (moving up in each panel). However, this relationship can be reversed at very high testing intensities (provided testing is targeted, and $\theta\_w$ is relatively small; \fref{pan2}(b), right three panels of top row). It is theoretically possible for increasing testing intensity to \emph{increase} \Rnum because more rapid testing leaves more susceptibles in the ``waiting-for-negative-results'' category at the DFE; if these people become infected while waiting, they will need to wait for their negative test result before they can be tested again, receive a positive test, and then begin self-isolating. This effect is usually weak compared to the beneficial effects of testing.

\end{itemize}

\section{Discussion}

In this paper, we have developed and analyzed a simple compartmental model that combines epidemiological dynamics---as defined by a simple SIR model---with the dynamics of testing and isolation. Our model is a caricature: it models the most basic feedbacks between epidemic and testing processes, but does not attempt to incorporate the many known complications of COVID-19 epidemiology (e.g., exposed, pre-symptomatic, and asymptomatic compartments \citep{kain2021chopping}; time-varying testing rates; behavioural dynamics \citep{weitz2020awareness}). Thus, it is most appropriate for assessing the \emph{qualitative} phenomena that arise from the interactions between transmission dynamics and testing, rather than for making quantitative predictions or guiding pandemic responses.

Many of the qualitative results we have derived confirm simple, common-sense intuitions. In particular, we can generally decrease $\Rnum$ by increasing isolation efficacy or testing intensity; returning tests faster, if individuals do not isolate while they are waiting for results; or increasing testing focus to target individuals who are likely to be infectious (e.g., symptomatic people or close contacts of known infections).

However, we did find two surprising patterns: under some conditions longer delays in returning tests can reduce epidemic spread, and increasing testing rates can increase spread.

Over broad regions of parameter space, decreasing $\omega$---i.e., \emph{slowing} the rate at which test results are returned---\emph{decreases} $\Rnum$ (for random testing, the parameter region is $\theta\_w \gtrsim 0.25$; for targeted testing, $\theta\_w \geq 0.25$ and either $\theta\_c \ge 0.5$ or $1/\omega > 5$; see Table~\ref{tab:params} for parameter definitions).  This result is counterintuitive and would not be expected by public health authorities that have invested a great deal of effort in reducing delays from testing to results. Dynamically, this effect occurs because speeding up test returns shortens the isolation period of uninfected individuals (for infected people it only shortens the time to progression to the isolation level of the confirmed-positive compartment). Slowing test returns increases $\Rnum$ only if the proportion of infectives in the tested population is high and isolation is relatively strong among people waiting for test results. 

While slowing test returns does decrease $\Rnum$ over broad regions of parameter space in our model, there are several real-world processes missing from our model that make it unlikely that slowing test returns would actually be an effective public health measure. First, we do not model the primary benefit of rapid testing, i.e., detecting and containing outbreaks while they are still in progress. This process could be modeled phenomenologically by making the testing focus more targeted as an increasing proportion of cases is detected, because finding infections allows tests to be concentrated on their connections. Second, individuals may become less likely to maintain isolation if they are required to do so for longer; phenomenologically, we could allow  effectiveness of isolation in the waiting population to be an increasing function of test-return speed, or we could introduce a separate ``waiting, but no longer isolating'' compartment that individuals entered from the ``waiting, isolated'' compartment at a specified rate. Finally, if one wants to decrease the overall transmission rate of the population there are more effective methods than keeping tested people in limbo; these include masking, ventilation, distancing measures, retail and event closures, and stay-at-home orders.

The other counterintuitive result from our analysis is that, for sufficiently high testing intensity $\rho$, further increasing testing intensity can actually \emph{increase} $\Rnum$ (e.g., \fref{pan2}(b), upper right panel [$\theta\_c=1$, $\theta\_w=0$]). This phenomenon can occur because we are considering the DFE in the presence of testing; thus there is an equilibrium distribution of susceptibles between the $S\_n$ (untested) and $S\_u$ (waiting) compartments even as the disease approaches extinction. A higher rate of testing leads to a greater proportion of individuals waiting for negative tests at the DFE. If infected, individuals in this group will take longer to be tested again and to subsequently isolate (because they must wait for their negative tests to be returned before being tested again). If isolation in this group ($\theta\_w$) is low, this effect can under some (relatively rare) circumstances (high $\theta\_c$, low $\omega$, high $\rho$) allow $\Rnum$ to increase with testing intensity. 
We can show that this phenomenon occurs \emph{only} under targeted testing ($w_I > w_S$), but we have not yet found a simple explanation of why it cannot occur under random (unfocused) testing. This phenomenon is also unlikely to occur in the real world. In particular, it depends on levels of testing that are unrealistically high (at least in large, general-population settings).

Although we model the testing process in more detail than typical epidemiological models, one place where more detail could be informative is in the processes determining the testing weights $\{w_S, w_I, w_R\}$. While random testing, as done for surveillance purposes, unambiguously leads to equal testing weights, making precise quantitative connections between public-health practices and testing weights is difficult in other contexts. The testing weights reflect the correlation between an individual's risk of infection and their likelihood of being tested due to age, occupation, geographic location, etc.. This correlation is influenced, among many other factors, by the proportion of the uninfected population with COVID-like symptoms (e.g., due to seasonal upper respiratory tract infections); the concentration of transmission and testing in hot spots such as long-term care facilities and high-density workplaces; the overall testing intensity (and hence, e.g., restriction to symptomatic individuals); and the proportion of COVID-infected people who are symptomatic. Future steps should explore mathematically tractable ways to model some of these factors more precisely. For example, separating the infected class into exposed, symptomatic, and a- or pre-symptomatic compartments and allowing the testing weights to vary across non-symptomatic (exposed/asymptomatic/presymptomatic) vs.\ symptomatic compartments could reflect the allocation of tests for diagnostic purposes (targeting symptomatic individuals) vs.\ contact-tracing (targeting infected but non-symptomatic individuals) vs.\ screening (relatively equal weights, depending on the venue). Alternatively, one could make the testing weights depend on the testing intensity or test-return rate as suggested above. Whatever complexity is added would probably put the model beyond reach of the analytical methods we have used in this paper, but one could still use semi-numerical methods such as constructing the next-generation matrix and using it to evaluate the derivatives of $\Rnum$ with respect to the parameters numerically.

Although testing and tracing is a key part of infection control strategies, mathematical epidemiologists have typically analyzed it with detailed models designed to inform particular public health efforts \citep{endo2020implication,hellewell2020feasibility,jenness2020modeling}, rather than analyzing simple but general models of the feedback between testing and transmission dynamics.
There have been several modeling studies of testing and tracing dynamics and their interaction with epidemiological dynamics. In the context of repeated screening and random testing of isolated populations (such as the members of a university), \cite{bergstrom2020frequency} provided analytical results quantifying the effects that proactive screening of asymptomatic individuals and isolation of confirmed-positive cases could have in reducing the spread of disease. 
\cite{rogers2021high} simulated a SEIR model with testing and isolation; they similarly suggest a strategy of rapid testing with antigen tests and the subsequent isolation of confirmed-positive individuals.
\cite{friston2021testing} model the effects of self-isolation on testing and tracing with a focus on projections under different testing and tracing scenarios. They conclude that the emergence of a second wave depends almost primarily on the rate at which immunity is lost and that it is necessary to track asymptomatic individuals in order to control the outbreak. 
Our modeling approach differs from these previous efforts in that it examines the effects of test-return rates and of different levels of testing focus, from random to highly targeted. We hope this paper will inspire further explorations of the fundamental properties of epidemic models that incorporate explicit testing processes.

\bibliography{SIRlibrary}

\clearpage
\appendix
\section{Appendix}
\setcounter{equation}{0}
\setcounter{figure}{0}
\setcounter{table}{0}
\makeatletter
\renewcommand{\theequation}{A\arabic{equation}}
\renewcommand{\thefigure}{A\arabic{figure}}
\renewcommand{\bibnumfmt}[1]{[A#1]}
\renewcommand{\citenumfont}[1]{A#1}

\subsection{Model and calculation of $\Rnum$}\label{app:R0}

The model in the form of a system of ordinary differential equations is 
\begin{subequations}\label{model}
\begin{align}
 d S\_u/dt &= -\Lambda S\_u - \testing{S} S\_u + \omega S\_n, \\
 d S\_n/dt &= -(1-\theta\_w)\Lambda S\_n + (1-p_S) \testing{S} S\_u - \omega S\_n, \\
 d S\_p/dt &= -(1-\theta\_w)\Lambda S\_p + p_S \testing{S} S\_u - \omega S\_p, \\
 d S\_c/dt &= -(1-\theta\_c)\Lambda S\_c + \omega S\_p, \\
 d I\_u/dt &= \Lambda S\_u - \testing{I} I\_u + \omega I\_n  - \gamma I\_u, \\
 d I\_n/dt &= (1-\theta\_w)\Lambda S\_n + (1-p_I) \testing{I} I\_u - \omega I\_n -\gamma I\_n, \\
 d I\_p/dt &= (1-\theta\_w)\Lambda S\_p + p_I \testing{I} I\_u - \omega I\_p -\gamma I\_p, \\
 d I\_c/dt &= (1-\theta\_c)\Lambda S\_c + \omega I\_p - \gamma I\_c, \\
 d R\_u/dt &= \gamma I\_u - \testing{R} R\_u + \omega R\_n, \\
 d R\_n/dt &= \gamma I\_n + (1-p_R) \testing{R} R\_u - \omega R\_n, \\
 d R\_p/dt &= \gamma I\_p + p_R \testing{R} R\_u  - \omega R\_p, \\
 d R\_c/dt&= \gamma I\_c + \omega R\_p, \\
 dN/dt &= \omega (S\_n + I\_n + R\_n),  \\
 dP/dt &= \omega(I\_p + R\_p) ,
\end{align}
\end{subequations}
(see Table \ref{tab:params} for parameter definitions). The next generation matrix for this model is $G = F V^{-1}$, where matrix $F$ represents the inflow of new infection to the infected compartments and matrix $V$ represents the flow in the infected compartments when the population is totally susceptible. 
Matrices $F$ and $V$ are
\begin{align}
\label{F}
F =& \beta/N \left[ \begin {array}{cccc} 
S\_u^* & (1-\theta\_w) S\_u^* & (1-\theta\_w) S\_u^* & (1-\theta\_c) S\_u^*\\
(1-\theta\_w)S\_n^* & (1-\theta\_w)^2 S\_n^* & (1-\theta\_w)^2 S\_n^* & (1-\theta\_w)(1-\theta\_c) S\_n^* \\ 
0&0&0&0\\
0&0&0&0
 \end {array} \right] \\ \notag
 =&\beta/N \left[ \begin {array}{c} S\_u^* \\ (1-\theta\_w)S\_n^* \\ 0 \\ 0 \end {array} \right]
        \left[ \begin {array}{cccc} 1,   1-\theta\_w,   1-\theta\_w,   1-\theta\_c \end {array} \right],
\end{align}
and 
\begin{align}
\label{V}
V= \left[ \begin {array}{cccc}  
\testinghat{I}+\gamma&-\omega&0&0\\
-(1-p_I)\testinghat{I}&\omega+\gamma&0&0\\
-p_I \testinghat{I}&0&\omega+\gamma&0\\
0&0&-\omega&\gamma
\end {array} \right].
\end{align}
The matrix inverse of $V$ is 
\begin{align}
\label{vinv}
V^{-1} =
\frac{1}{\gamma C}
\left[ \begin {array}{cccc}
\gamma(\omega+\gamma)^2 & \gamma\omega(\omega+\gamma)&0&0\\ \noalign{\medskip}
\gamma(\omega+\gamma)(1-p_I) \testinghat{I} & \gamma(\omega+\gamma)(\testinghat{I}+\gamma)&0&0 \\ \noalign{\medskip}
\gamma(\omega+\gamma) p_I \testinghat{I} & \gamma\omega p_I \testinghat{I} & C \gamma/(\omega+\gamma) & 0 \\ \noalign{\medskip}
\omega(\omega+\gamma) p_I \testinghat{I} & \omega^2 p_I \testinghat{I} & C \omega/(\omega+\gamma) & C
\end {array} \right],
\end{align}
where $C= \big( \gamma(\omega+\gamma)+(\gamma+\omega p_I)\testinghat{I} \big) (\omega+\gamma)$ and $\testinghat{I}$ is the \percap testing rate for the infected people and represented in Eq.~\eqref{eq:fi}. Note that all the columns of matrix $V^{-1}$ summ up to $1/\gamma$.

The particular form of $F$ with two rows of zeros at the bottom results in the following blocked form of matrix $G$.
\begin{equation}
\label{mat:G}
G = \left[ \begin {array}{cc}
G_{11}&G_{12}\\
0&0
\end {array} \right],
\end{equation}
where both blocked matrices $G_{11}$ and $G_{12}$ are 2 by 2. Given the upper triangular form of matrix $G$, the basic reproduction number $\Rnum$ (defined as the spectral radius of matrix $G$) is only determined by the blocked matrix $G_{11}$,
\begin{equation}
\label{G11}
G_{11} = \frac{\beta}{\gamma   C} 
\left[ \begin {array}{c} (\omega-\rho)/\omega \\ (1-\theta\_w)\rho/\omega \end {array} \right]
\left[ \begin {array}{cccc} 1,   1-\theta\_w,   1-\theta\_w,   1-\theta\_c \end {array} \right]
\left[ \begin {array}{cc}
\gamma(\omega+\gamma)^2 & \gamma\omega(\omega+\gamma)\\ \noalign{\medskip}
\gamma(\omega+\gamma)(1-p_I)\testinghat{I} & \gamma(\omega+\gamma)(\testinghat{I}+\gamma) \\ \noalign{\medskip}
\gamma(\omega+\gamma) p_I \testinghat{I} & \gamma\omega p_I \testinghat{I} \\ \noalign{\medskip}
\omega(\omega+\gamma) p_I \testinghat{I} & \omega^2 p_I \testinghat{I}
\end {array} \right].
\end{equation}
It is notable that matrix $F$ \eqref{F} has rank one and consequently $G_{11}$ does so. That is $G_{11}$ has only one non-zero eigenvalue which is $\Rnum$.

The expression of $\Rnum$ has a complicated form with all of the model parameters involved. This expression can be simplified and represented given the specific form of matrix $G_{11}$ \eqref{G11}. For the purpose of simplicity we present $\Rnum$ in the manuscript in terms of expressions $C$, $C_1$ and $C_2$, specified in \eqref{eq:C12}. 

It remains difficult to show that the reproduction number $\Rnum$ is decreasing with respect to \percap testing intensity, $\rho$, and the speed of the test return, $\omega$, for the feasible ranges of the parameters, that is
\begin{align}
\label{cond}
& \omega>0, \\
& 0 \leq \rho<\omega,\\ 
& 0 \leq \theta\_w \leq \theta\_c \leq 1, \\
& \oldtesttarget\geq 1.
\end{align}
In realistic cases the testing rate $\rho$ is very small (i.e., only a small fraction of the population can be tested every day); it is thus reasonable to use a linear approximation of $\Rnum$ for $\rho \ll 1$ to analyze the behaviour of $\Rnum$ with respect to $\omega$ (see section \appref{omega}). 
In the next section we provide an equivalent representation of $\Rnum$ in order to show that increasing testing intensity typically decreases $\Rnum$.

\subsection{More testing intensity may decrease $\Rnum$}\label{app:rho}

This section shows that $\pro{\Delta}$ can be positive or negative, with $\Delta$ defined in Eq.~\eqref{eq:C12}, and thus $\pro{\Rnum} < 0$, where $\Rnum$ is given in Eq.~\eqref{R0}. We rewrite matrix $G_{11}$ in \eqref{G11} in the following form to simplify the calculations:
\begin{equation}
\label{G112}
G_{11} = \frac{\beta}{\gamma C} 
\left[ \begin {array}{c}  S\_u^*/N \\ (1-\theta\_w) S\_n^*/N  \end {array} \right]
\left[ \begin {array}{cccc} 
C-C_1, C-C_2\end {array} \right],
\end{equation}
where $C$ is the same as the one in Eq. \eqref{eq:C12}, i.e., 
$$C=(\omega+\gamma)(\gamma(\omega+\gamma)+(\omega p_I+\gamma) \testinghat{I}),$$
and $C_1$ and $C_2$ are 
\begin{align*}
C_1 =& (\omega+\gamma)(\theta\_w   \gamma+\theta\_c  \omega  p_I) \testinghat{I},\\
C_2 =& \big( \omega\gamma(1+p_I)\testinghat{I}+\gamma^2 (\omega+\gamma+\testinghat{I})\big)\theta\_w + \omega^2 p_I \testinghat{I} \theta\_c,
\end{align*}
where $\testinghat{I}$ is given in Eq.~\eqref{eq:fi}.
Note that for analysis brevity, we let $N=1$, thus $S\_u^*$ and $S\_n^*$ are in the scale of 0 to 1.
$\Rnum$ is in the same form as in Eq. \eqref{R0}  
$$\Rnum= \frac{\beta}{\gamma} (1-\Delta),$$
where 
\begin{equation*}
\label{eq:del4}
\Delta= \frac{1}{C}\big(C_1 S\_u^*+(C_2(1-\theta\_w)+C \theta\_w) S\_n^*\big).
\end{equation*}

The first goal
is to explore how changes in isolation, $\theta\_w$ and $\theta\_c$, affects $\Rnum$. Mathematically we would like to verify the sign of $\pder \Rnum{\theta\_w}$ and $\pder \Rnum{\theta\_c}$. We start with simplifying $\Delta$ \eqref{eq:del4} by factoring $\theta\_w$ and $\theta\_c$ in Eq.~\eqref{eq:del4}. Thus, $\Delta$ can be rewritten as
\begin{equation}
\label{eq:del4_theta}
\Delta= \frac{1}{C}\Big(
-D_1 S\_n^* \theta\_w^2 +
\big(-\omega^2 p_I \testinghat{I} S\_n^*\theta\_c+ D_2 S\_n^*+\gamma \testinghat{I}(\omega+\gamma) \big)\theta\_w+
(\omega+\gamma S\_u^*)\omega p_I \testinghat{I} \theta\_c
\Big),
\end{equation}
where
\begin{align}
\label{eq:del4_d1d2}
D_1=& (\omega+\gamma)\gamma^2+(\omega+\gamma+\omega p_I)\gamma \testinghat{I}, \\
D_2=& (3\omega+2\gamma)\gamma^2+(\omega+\gamma+2\omega p_I)\gamma \testinghat{I}+(\gamma+p_I \testinghat{I})\omega^2.
\end{align}
$\Delta$, Eq.~\eqref{eq:del4_theta}, is linear in $\theta\_c$ with a positive coefficient. thus
\begin{equation}
\label{eq:d_del4_thetac}
\pder\Delta{\theta\_c}=1/C(\gamma S\_u^*+\omega(1-\theta\_w S\_n^*))\omega p_I \testinghat{I}.
\end{equation}
This results in increasing $\Delta$, thus decreasing $\Rnum$ with respect to $\theta\_c$, that is $\pder \Rnum{\theta\_c}\leq 0$. Note that $C$ is independent of $\theta\_c$ and $\theta\_w$. 

With a similar logic, $\Delta$ \eqref{eq:del4_theta} is a concave-down quadratic equation in $\theta\_w$, given by
\begin{equation}
\label{eq:del4_thetaw}
1/C \Big( -D_1 S\_n^*\theta\_w^2+
\big(-\omega^2 p_I \testinghat{I} S\_n^*\theta\_c+D_2 S\_n^*+\gamma \testinghat{I}(\omega+\gamma) \big)\theta\_w\Big).
\end{equation}
We show that the feasible range of $\theta\_w$ lies between 0 and the vertex of this parabola where the parabola is increasing in $\theta\_w$, and so does $\Delta$ which results in inferring $\pder\Rnum{\theta\_w}\leq 0$.
It is enough to show that partial derivative of the expression \eqref{eq:del4_thetaw} with respect to $\theta\_w$ at $\theta\_w=1$ is non-negative. It follows that
\begin{align}
\label{eq:d_del4_thetaw}
\pder\Delta{\theta\_w}\bigg\rvert_{\theta\_w=1} =&
1/C\Big( (D_2-2D_1-\omega^2 p_I \testinghat{I} \theta\_c) S\_n^*+\gamma \testinghat{I}(\omega+\gamma) \Big) \\\notag
=& 1/C\Big( (\gamma(\omega+\gamma)+\gamma\omega^2+(1-\theta\_c)\omega^2 p_I \testinghat{I} ) S\_n^*
+\gamma(\omega+\gamma) \testinghat{I} (1-S\_n^*) \Big),
\end{align}
which is a positive quantity, given that $\theta\_c$ and $S\_n^*$ vary between 0 and 1.

The second goal is to explore how changes in \percap testing intensity $\rho$ affects $\Rnum$. Mathematically we would like to verify the sign of $\pder\Rnum{\rho}$, which specifically depends on $\pder\Delta{\rho}$. We use the derived expressions for $S\_u^*$ and $S\_n^*$, given by Eqs.~\eqref{dfe}, in $\Delta$ \eqref{eq:del4}. Also, we define $\phi = \testinghat{S} = \frac{\rho \omega}{\omega-\rho}$, to reparameterize $\rho$. This is mainly to avoid singularity in $\testinghat{I}$ \eqref{eq:fi}, when testing intensity $\rho$ is very close to the rate of test return $\omega$. Thus, $\rho$ is reparameterized as 
\begin{equation}
\label{eq:phi}
\rho=\frac{\omega \phi}{\omega+\phi}.
\end{equation}
This one-to-one monotonic reparameterization enables us to simplify the mathematical expressions and explore the simpler $\pder\Delta{\phi}$ instead of the complicated $\pder\Delta{\rho}$.
Defining $\testtarget \equiv \oldtesttarget$, the derivative is 
\begin{align}
\label{eq:dd3dr}
\partial\Delta/\partial\phi=& \frac{1}{d_3} (a_3 \phi^2+b_3 \phi+c_3),
\end{align}
where
\begin{align}
\label{eq:abcd2}
a_3& = \testtarget \Bigg(
&& (1-\theta\_w) (1+\testtarget)\theta\_w \gamma^3 +(1-\theta\_c) p_I^2 \theta\_w \testtarget \omega^3 \\ \notag
& ~ &&+\Big( \big( (1-\theta\_w-\testtarget) \theta\_c +(3-2\theta\_w) \theta\_w\testtarget \big) p_I +(1-\theta\_w) (1+\testtarget)\theta\_w  \Big) \omega \gamma^2 \\ \notag
& ~ &&+\Big(
\big((1-\theta\_w-\theta\_w \testtarget) \theta\_c+ (2-\theta\_w) \theta\_w \testtarget \big) p_I
+(2 \theta\_w-\theta\_w^2-\theta\_c) \testtarget p_I^2
\Big) \omega^2 \gamma \Bigg), \\ \notag
b_3& = && \hspace{-1cm} 2\testtarget (\omega+\gamma)\gamma\Big(
(\omega+\gamma+\omega p_I)(2-\theta\_w) \gamma \theta\_w+ (1-\theta\_w) \omega^2 p_I\theta\_c+\omega^2 p_I\theta\_w
\Big) ,\\ \notag
c_3& = && \hspace{-1cm} (\omega+\gamma)^2 \gamma\Big(
(2-\theta\_w) \gamma^2 \theta\_w+
(1+\testtarget) \omega \gamma \theta\_w+
\testtarget \omega^2 p_I\theta\_c \Big),\\ \notag
d_3&  = && \hspace{-1cm} \frac{(\omega+\gamma)}{\omega}  \Big((\omega p_I+\gamma)\testtarget\phi + (\omega+\gamma)\gamma \Big)^2(\omega+\phi)^2.
\end{align}
Note that $\phi\geq 0$, also $b_3$, $c_3$ and $d_3$ are all positive. However $a_3$ can be positive or negative.
If $a_3\geq 0$, $\partial\Delta/\partial\phi \geq 0$ for all feasible range of parameters, thus $\pro\Rnum \leq 0$. It is straightforward to show that $a_3\geq 0$ when testing is random, i.e., $w\_S=w\_I=1$. 
If $a_3 < 0$, then the quadratic expression in the numerator of \eqref{eq:dd3dr} has a positive root, $\phi^*$, such that for $\phi>\phi^*$, $\partial\Delta/\partial\phi < 0$. 

An example of this countervailing effect of $\phi$, and consequently $\rho$, on $\Rnum$ occurs when $\theta\_w=0$ and $\theta\_c=1$.
This is illustrated in the top-right panel of the \fref{pan2} panel (b), where the strength of isolation for awaiting people is the least, but the most for the confirmed cases. In this case, simplifying $a_3$ in Eq.~\eqref{eq:abcd2} gives
\begin{equation}
  \begin{split}
    a_3 & =\testtarget \omega \gamma p_I\left((\omega+\gamma)-\testtarget(\omega p_I+\gamma)\right)  \\
    & \propto  \omega \left( 1- \testtarget p_I \right) + \gamma \left( 1- \testtarget \right)  .
  \end{split}
\end{equation}
If $p_I>0$, then $a_3<0$ for sufficiently targeted testing (i.e. when $\testtarget p_I >1$; because
$p_I \leq 1$, \testtarget is always $\geq \testtarget p_I$).
When the test is perfectly sensitive ($p_I=1$), $a_3<0$ as long as $\testtarget>1$.
Under either of these conditions, there exists a range for $\rho$ over which $\pder\Rnum{\rho}\leq 0$.  
Because increasing values of $\rho$ and $\omega$ both delay the rate at which individuals flow to the $I\_c$ compartment, it is reasonable that increasing either value could (under appropriate circumstances) increase $\Rnum$.

\subsection{Rate of returning tests} \label{app:omega}
The third goal is to explore how changes in the rate of test return $\omega$ affect $\Rnum$. Mathematically we would like to verify the sign of $\pder\Rnum{\omega}$, which specifically depends on $\pder\Delta{\omega}$. We use
the linearization of $\Delta$ when $\rho \ll 1$ to show that there a non-monotonic relationship between $\Rnum$  and $\omega$. The linear term in the Taylor expansion of $\Delta$ when $\rho \ll 1$ is
\begin{equation}
\label{eq:lin}
\Delta = \frac{\rho}{\omega \gamma(\omega+\gamma)} \Big(
\testtarget \omega^2 p_I \theta\_c+(\testtarget+1) \gamma\omega\theta\_w +\gamma^2 \theta\_w(2-\theta\_w)
\Big). 
\end{equation}
This results in
\begin{equation}
\label{eq:dlindo}
\pder\Delta{\omega} = \frac{\rho}{\omega^2 (\omega+\gamma)^2} \Big(
(p_I \testtarget\theta\_c -(1+\testtarget)\theta\_w)\omega^2-2\theta\_w \gamma(2-\theta\_w) \omega+\theta\_w \gamma^2 (\theta\_w-2)
\Big).
\end{equation}
The latter expression has two roots
\begin{align}
\label{eq:omega_roots}
\omega^*_{+} &= \frac{\gamma\Big(-c_4 + \sqrt{c_4^2 + (a_4-b_4)c_4}\Big)}{b_4-a_4} \\
\omega^*_{-} &= \frac{\gamma\Big(-c_4 - \sqrt{c_4^2 + (a_4-b_4)c_4}\Big)}{b_4-a_4},
\end{align}
where $a_4 = p_I\testtarget\theta\_c$, $b_4 = (1+\testtarget)\theta\_w$, $c_4 = \theta\_w(2-\theta\_w)$. This enables us to describe the behaviour of $\Rnum$ with respect to $\omega$ in the following two cases.
\begin{itemize}
\item \textbf{Case I:} If $b_4 \geq a_4$, $\pder\Delta{\omega} < 0$, so $\Rnum$ is always increasing with respect to $\omega$ (i.e., it is always harmful to return tests more rapidly). 
\item \textbf{Case II:} If $b_4 < a_4$, $\Rnum$ will be decreasing with respect to $\omega$ (i.e., returning tests more rapidly is beneficial) only when $\omega > \omega^*_-.$
\end{itemize}
Note that $b_4 \geq a_4$ is characterized by
\begin{equation}\label{case_1_ineq}
\Big(\testtarget+1\Big)\theta\_w \geq \testtarget p_I\theta\_c \Longleftrightarrow \Big(\frac{w_S}{w_I}+1\Big)\frac{1}{p_I} \geq \frac{\theta\_c}{\theta\_w}.
\end{equation}
We begin with a \textbf{proof of Case I}. Suppose that $b_4 \geq a_4$. If the roots of $\pder\Delta{\omega}$ are not complex, then we must have $c_4 \geq b_4-a_4$. Note that $\omega^*_{-}$ must be negative since the numerator is clearly negative but the denominator is positive. Next, note that since $c_4 \geq b_4-a_4$, the numerator of $\omega^*_{+}$ must also be negative, so $\omega^*_{+}$ is negative. Thus, in this case, $\pder\Rnum{\omega}$ does not change sign on $(0,\infty)$. Checking the sign of \ref{eq:dlindo} for arbitrarily large $\omega$ shows that it is negative (since the $(a_4-b_4)\omega^2$ term dominates and is negative). So $\Rnum$ is increasing with respect to $\omega$ on all of $(0,\infty)$.

We now turn our attention to a \textbf{proof of Case II}. Suppose that $b_4 < a_4$. It follows that $\omega^*_+$ and $\omega^*_-$ are real since $c_4 > 0 > b_4-a_4$. Next, note that $\omega^*_-$ is positive since both the numerator and denominator are negative. On the other hand, $\omega^*_+$ is negative since the denominator is negative but the numerator is positive (because $c_4 > b_4-a_4$). Thus, our task is to understand the sign of $\pder\Delta{\omega}$ around the root $\omega^*_-$. Checking the sign of \ref{eq:dlindo} for arbitrarily large $\omega$ shows that it is positive (since the $(a_4-b_4)\omega^2$ term dominates and is positive). Likewise, checking the sign for values of $\omega$ close to $0$ shows that it is negative. Thus, $\Rnum$ is increasing with respect to $\omega$ when $\omega < \omega^*_-$, and is decreasing $\omega > \omega^*_-$. 

Having presented the formal analysis, we now concern ourselves with its biological interpretations. We begin by interpreting \ref{case_1_ineq}, under which returning tests more rapidly is \emph{always} harmful. Notice that the ratio $\frac{\theta\_c}{\theta\_w}$ is simply a measure of how much more strongly individuals self-isolate when they test positive compared to when waiting for tests. Since the rate of test return directly influences the rate at which individuals change from a waiting state to a confirmed-positive state, it is intuitive that $\frac{\theta\_c}{\theta\_w}$ would appear in \ref{case_1_ineq}. Next, note that the left-hand side increases when test sensitivity decreases and when targeting of positive individuals is poor. This is consistent with our intuition: a false negative that is returned more rapidly will allow an infectious individual to relax their self-isolation, thus increasing transmission. Likewise, if individuals tested are mainly susceptible (rather than infectious), then returning tests more slowly would encourage them to self-isolate for longer while awaiting test results. Having understood the role of each of the parameters in \ref{case_1_ineq}, a holistic interpretation of this inequality is that returning tests more slowly is helpful when the benefit of extended self-isolation by infected individuals awaiting test results outweighs the benefit of identifying positive cases. 

Now we interpret \textbf{Case II}.  In this case, $\Rnum$ will have a global maximum with respect to $\omega$ at $\omega^*_-$. Note that our model assumption that $\rho < \omega$ plays an important role here: if $\omega^*_- < \rho$, then $\Rnum$ will be always decreasing with respect to $\omega$. On the other hand, if $\rho < \omega^*_-$, then $\Rnum$ will be increasing with respect to $\omega$ on $(\rho, \omega^*_-)$ and decreasing beyond that.

\subsection{The effect of testing focus parameter $\testtarget$ on $\Rnum$} \label{app:w}

\begin{align}
\label{eq:d_del_wis}
\pder \Delta{w\_{IS}}= \dfrac{(\omega-\rho) (\omega(\omega-\rho\theta\_w)+\gamma(\omega-\rho)) (\theta\_w\gamma+\theta\_c\omega p\_I)}
{(-\omega^2 \gamma+\omega \gamma \rho-\gamma \rho \omega w\_{IS}-
\omega \gamma^2+\gamma^2 \rho-\omega^2 p_I \rho w\_{IS})^2},
\end{align}
which is a positive quantity. Thus, $\pder \Rnum{w\_{IS}}\leq 0$. Therefore, increasing the focus of testing on the infectious people will result in less transmission. 

\subsection{On Testing Rate and Numerical Singularity} \label{app:singularity}

In this work, we do not present any numerical solutions of the ODEs to investigate typical trajectories.
However, attempting to do so quickly reveals a problem in the way the model is posed;
with the scaling of testing ($\sigma$) defined as in the body of the paper (Eq.~\eqref{sigma}), the population in the $S$ compartments appeared to blow up when the system is near the DFE. This occurs because once the only untested people are susceptibles, the FOI approaches $\Lambda=0$, and the testing rate $\testing{S} \to \rho N/S\_u$. Thus, the first equation of the model \eqref{model} will become
$d S\_u/dt = - \rho N + \omega S\_n$. Thus changes in $S\_u$ will be independent on $S\_u$, and the decay of the $S\_u$
population becomes linear rather than exponential --- allowing $S\_u$ to become negative!
To avoid this problem the testing rate, $\sigma$, should be formulated such that people from the untested compartments will not be tested if they are not there.
One way to fix this issue, is to consider a maximum testing rate, $\tau$ (1/day). In general, we want to test at a rate of $\rho$ across the whole population. This won't always be possible, so we impose a maximum rate of $\tau$ per testable person and redefine $\sigma = \frac{\tau \rho N}{\tau W + \rho N}$, with the assumption that $\tau \gg \rho$. This modification of $\sigma$ does not affect any of the results we have derived about the invasion of the epidemic from the DFE (i.e., results on $\Rnum$ and $\Delta$).

\end{document}

%% file: modeldefs.tex
\newcommand{\betaparam}{0.5}

\newcommand{\invgammaparam}{6.0}
\newcommand{\Rnumparam}{3.0}
\newcommand{\rparam}{0.3}